\documentclass[twocolumn,prd,aps]{revtex4}

\usepackage{amsmath}
\usepackage{verbatim}
\usepackage{graphicx}
\usepackage[usenames]{color}
\usepackage{psfrag}
\usepackage[ps2pdf,colorlinks,bookmarks]{hyperref}

\definecolor{linkblue}{rgb}{0,0,0.8}
\definecolor{linkgreen}{rgb}{0,0.5,0}
\hypersetup{pdfpagemode=None, pdfstartview=FitH, linkcolor=linkblue, %
            citecolor=linkgreen, urlcolor=linkblue}

\def\beq{\begin{equation}}
\def\eeq{\end{equation}}
\def\bea{\setlength\arraycolsep{1.4pt}\begin{eqnarray}}
\def\eea{\end{eqnarray}}
\def\bit{\begin{itemize}}
\def\eit{\end{itemize}}

\def\ld{\left}
\def\rd{\right}

\def\fr{\frac}
\def\oo{\frac{1}}

\def\sig{\sigma}

\def\bra{\left\langle}
\def\ket{\right\rangle}
\def\cij{C_{ij}}

\begin{document}

\title{No evidence for anomalously low variance circles on the sky}
\author{Adam Moss} \email{adammoss@phas.ubc.ca}
\affiliation{Department of Physics \& Astronomy\\
University of British Columbia, Vancouver, BC, V6T 1Z1  Canada}
\author{Douglas Scott} \email{dscott@phas.ubc.ca}
\affiliation{Department of Physics \& Astronomy\\
University of British Columbia, Vancouver, BC, V6T 1Z1  Canada}
\author{James P. Zibin} \email{zibin@phas.ubc.ca}
\affiliation{Department of Physics \& Astronomy\\
University of British Columbia, Vancouver, BC, V6T 1Z1  Canada}

\date{\today}

\begin{abstract}
In a recent paper, Gurzadyan \& Penrose claim to have found directions on the sky centred on which are circles of anomalously low variance in the cosmic microwave background (CMB).  These features are presented as evidence for a particular picture of the very early Universe.  We attempted to repeat the analysis of these authors, and we can indeed confirm that such variations do exist in the temperature variance for annuli around points in the data. However, we find that this variation is entirely expected in a sky which contains the usual CMB anisotropies.  In other words, properly simulated Gaussian CMB data contain just the sorts of variations claimed. Gurzadyan \& Penrose have not found evidence for pre-Big Bang phenomena, but have simply re-discovered that the CMB contains structure.
\end{abstract}

\maketitle
%\noindent
%{\em Introduction.---}%
A recent preprint by Gurzadyan \& Penrose \cite{GurPen} (hereafter GP10) makes what must be considered by any standard to be an extraordinary claim -- that the cosmic microwave background (CMB) contains statistically unusual circles which are signatures of events which happened before the hot, dense stage in the usual Big Bang theory.  In particular, GP10 claim to have found directions on the sky centred on which are circles of anomalously low variance in the CMB.  If true, this would have profound implications for our understanding of the early Universe, since such features are not expected in the standard picture of primordial structure, based on Gaussian random fields.  Such a result would also be very surprising, given the intense scrutiny the CMB has seen in the search for signs of non-Gaussianity.  However, we will show that the evidence claimed in GP10 fails to be extraordinary; indeed, such low-variance circles are {\em expected\/} to arise, as properly simulated data show.

GP10 studied maps from the Wilkinson Microwave Anisotropy Probe ({\it WMAP\/}) 7-year data release \cite{Jarosik}.  They specifically looked at circles around particular positions, calculating the variance within annuli of $0.5^\circ$ width, and plotting this variance as a function of radius. They found some positions where there appeared to be concentric rings of suppressed variance.  In addition, figure~2 in GP10 implies that the scatter in the annulus variances is much larger in some particular locations than it is for a simulated CMB sky.

\begin{figure}
\centering
\includegraphics[width=\columnwidth,angle=0]{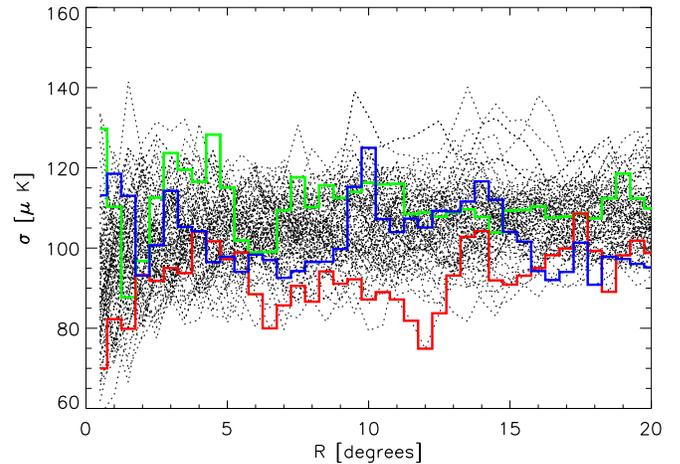}
\caption{\label{fig:Wband} Standard deviation in temperature measured within circular annuli bounded by radius $R$ and $R+dR$, with $dR = 0.5^\circ$, as a function of  $R$ for the {\it WMAP\/} 7-year W-band data.  The red and blue curves are for the two specific directions selected by GP10 (red corresponds to galactic coordinates $l = 105^\circ$, $b = 37^\circ$, and blue to $l = 252^\circ$, $b = -31^\circ$), and agree well with the results presented there.  The dashed curves are for 500 randomly chosen centres (only 100 are shown).  The green curve was chosen from the random centres on the basis that it has a prominent low-variance annulus at $R = 2.5^\circ$ as well as another near $R = 6^\circ$.  Other curves having annuli with high variance could just as easily be highlighted. Note also that the vertical axis on this plot does not start at zero, so the variations in variance are only at the ${\sim}\,$10\% level, and hence are quite subtle in the maps themselves.}
\end{figure}

We first tried to reproduce the plots of the variance in annuli presented for the two selected positions in GP10.  We found that we could reproduce very similar results (see Fig.~\ref{fig:Wband}), and that these are present in both W-band and V-band maps, as well as in the so-called ILC map (see \cite{Lambda}), which is largely free of foreground signals.  Hence it is clear that the particular map used is not the issue, and it is also unlikely that foregrounds are a major part of the effect.  Moreover, we found that strong variation among the annuli could be seen all over the map.  In other words, the particular positions selected in GP10 did not seem very unusual.  Additionally we could find rings of high variance just as easily as those of low variance (examples can be seen in Fig.~\ref{fig:Wband}); this is just what would be expected if both low and high variance rings are simply random excursions.

   For all of the results presented here we used the masked CMB maps. In each case we choose 500 random directions at Galactic latitudes of $|b| > 30^{\circ}$. Also, because of the {\it WMAP\/} scanning process, the number of samples going into individual pixels can vary, and so the uncertainty in the temperature varies across the sky.  Therefore our results include proper noise weighting
and removal of bias. As a check, we also performed the calculations using the {\em unmasked\/} sky, and not taking into account noise weighting.  In this case, our results for the two directions chosen in GP10 matched those of GP10 almost perfectly. However, ignoring these details did not have a large effect on our results (despite one of the directions suggested in GP10 being very close to the north ecliptic pole). 

\begin{figure}
\centering\includegraphics[width=\columnwidth,angle=0] {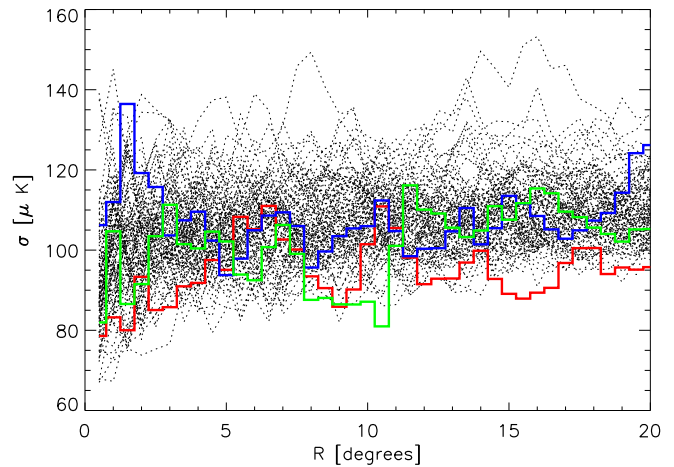} \caption{\label{fig:Wsim} Same as Fig.~\ref{fig:Wband}, except that this plot is for {\it simulated\/} W-band data.  We have plotted the results for 100 random directions.  The green and blue (heavy) curves were chosen from the full set of 500 random centres on the basis that they have annuli which show prominent dips in variance. The red (heavy) curve was also selected as it is systematically lower than average.   It is clear that there are directions around which there are low variance circles, and even families of concentric circles, just like in the real data.  Again, high variance annuli could just as easily be highlighted.}
\end{figure}

We next turned our attention to simulations.  We performed the same variance calculations as described above, but applied to a combination of simulated maps for all W-band feedhorns over the full 7 years (as provided on the LAMBDA website \cite{Lambda}); this should represent a realistic simulation of the CMB signal plus the {\it WMAP\/} noise. Our results are presented in Fig.~\ref{fig:Wsim}.  Comparing with Fig.~\ref{fig:Wband}, it is clear that the expectation of the annulus variances, together with their degree of fluctuation, are indistinguishable between the simulated data and the real data.  Since the {\it WMAP\/} data are dominated by the CMB anisotropies rather than by instrumental noise, it is not surprising that we found qualitatively the same results using simpler simulations which were based purely on the estimated power spectrum, $C_\ell$ \cite{Larson}, or equivalently the temperature angular correlation function, $C(\theta)$.

In Fig.~\ref{fig:Wsim} we also indicate specific curves which we selected based on the presence of particularly low variance annuli.  This demonstrates that the simulated data contain similar low-variance circles as occur in the real data (e.g.\ the green curve could be claimed to have a `family' of multiple low variance rings). The red curve was also chosen on the basis that it is systematically lower than average, similar to the GP10 $l = 105^\circ$, $b = 37^\circ$ direction. This example was found by searching 500 random directions, as opposed to 10885 in GP10. Hence it is clear that there is nothing unusual about this GP10 direction, and that it is simply an outlier with lower than average variance.

Results for a simulated sky are also presented in GP10 (the third panel of their figure~2), where the fluctuations in the variance appear very small compared with the real data.  Details of this simulation are not provided in GP10.  However, we found that we could reproduce a similar plot by using {\it purely white noise\/} simulations -- i.e.\ $C_\ell={\rm constant}$, normalized to have approximately the correct total power. It is clear that the scatter around the average variance level for the GP10 simulation is {\it not\/} what is expected in a map containing the observed CMB anisotropies; instead, it agrees with that for a spectrum with entirely the wrong scale dependence.

\begin{figure}
\centering
\includegraphics[width=\columnwidth,angle=0]{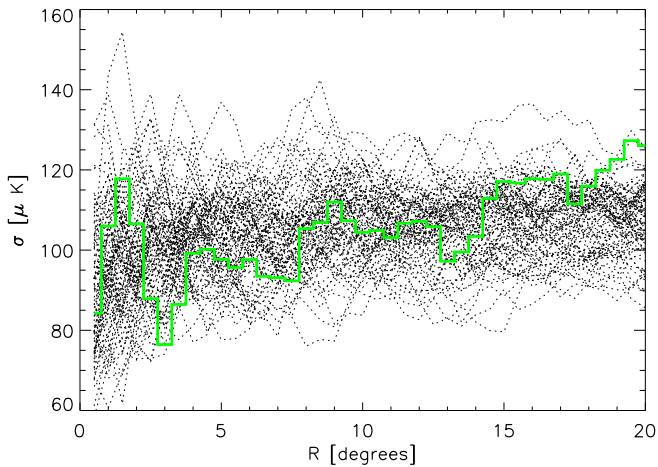}
\caption{\label{fig:Wtri} Same as Fig.~\ref{fig:Wband}, except that here we have calculated the variances in equilateral `triangular annuli' inscribed into circles of radius $R$.  We have plotted the results for 100 random directions and random orientations of the triangles.  The green (heavy) curve was chosen from the random centres on the basis that it has several prominent low-variance triangles.}
\end{figure}

It is important to point out that even when a low-variance annulus is identified in real or simulated data, that does not imply the presence of a circular feature in the map.  The low variance could simply be due to particularly low fluctuations in a {\em segment\/} of the annulus.  Indeed, it is straightforward to repeat the above analysis, but searching for low-variance features of shapes {\em other\/} than circular annuli. As an example, we performed the identical analysis steps, but looking instead for low-variance `triangular annuli', i.e.\ the regions between concentric equilateral triangles of different sizes \cite{triangles}.  The results are presented in Fig.~\ref{fig:Wtri}.  It is clear from this figure that there are directions around which there are similarly low-variance triangles to the low-variance circles found above.  Therefore there is nothing special about the presence of low-variance circles on the sky.  We also tested the same triangles in simulated data, finding results consistent with those in Fig.~\ref{fig:Wtri}.

We have refrained from any attempt to estimate the significance level for the specific patterns identified in GP10.  Since the predictions for low variance annuli are quite vague, it would be impossible to avoid the issue of a posteriori statistics.  But in any case, this is unnecessary, since it is clear that such low variance patterns occur by chance in realistically simulated skies.

Recalling that the CMB `signal' behaves statistically like a random `noise', and that it is Gaussian in nature, it is possible to understand theoretically the behaviour of the variances in the annuli.  All that is required is the $C_\ell$ power spectrum, or its real-space equivalent, the 2-point angular correlation function, which is essentially the variance on the sky as a function of angular separation between positions.  The variance of the temperature $T_i$ within an annulus, where $i$ labels the pixel, is \beq \sig^2_T = \oo{N}\sum_iT_i^2 - \oo{N^2}\sum_{ij}T_iT_j, \label{sig2} \eeq where $N$ is the number of pixels.  This is the variance for {\em one particular} realization of the Gaussian random field.  Taking the expectation over all realizations in the ensemble (denoted $\bra\cdots\ket$) gives \beq \bra\sig^2_T\ket = C(0) - \oo{N^2}\sum_{ij}\cij, \label{expsig2} \eeq where \beq \cij \equiv C(\theta_{ij}) = \bra T_iT_j\ket. \eeq In the limit of large $R$, we have $\bra\sig^2_T\ket \rightarrow C(0)$, i.e.\ the expected variance is independent of the shape of the region, and is given simply by the zero-lag angular correlation function $C(0)=\sum_\ell (2\ell+1)C_\ell/(4\pi)$. This is around $[ 110\,\mu {\rm K}]^2$, which we estimate from the actual WMAP W-band map, which includes noise. This matches the levels seen in Figs.~\ref{fig:Wband}--\ref{fig:Wtri} for large $R$.

\begin{figure}
\centering
\includegraphics[width=\columnwidth,angle=0]
{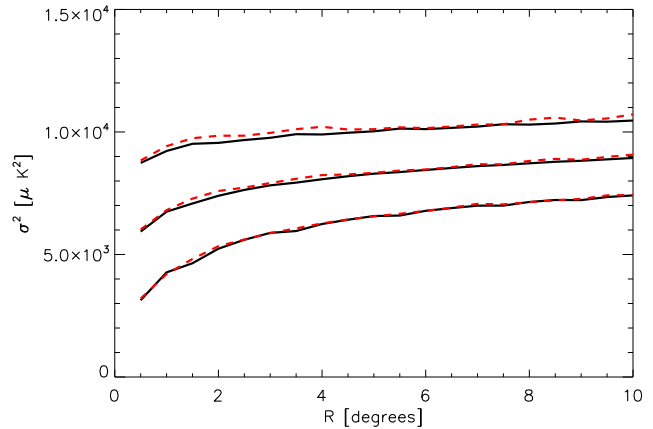} \caption{\label{fig:theory}  Theoretical values for the expectation (and $\pm 1 \sigma$ limits)  of the circular annulus {\em variance} (solid curves) compared to the mean and standard deviation from 1000 randomly chosen centres (dashed curves).  To compute these we used a {\em noiseless} full-sky map simulated from the best-fit {\it WMAP\/} 7-year power spectrum~\cite{Larson}. Note that in the previous figures we show the standard deviation of the temperature in each annulus, as opposed to the variance.}
\end{figure}

   Now we want the variance of $\sig^2_T$ over the ensemble.  So we want
\beq \bra\ld(\sig^2_T - \bra\sig^2_T\ket\rd)^2\ket = \bra\sig^4_T\ket 
                                              - \bra\sig^2_T\ket^2.
\eeq Straightforward computations give \bea \bra\ld(\sig^2_T - \bra\sig^2_T\ket\rd)^2\ket
  &=& \fr{2}{N^2}\sum_{ij}\cij^2 - \fr{4}{N^3}\sum_{ijk}\cij C_{ik}\nonumber\\
  &+& \fr{2}{N^4}\sum_{ijkl}\cij C_{kl}.
\label{varvar} \eea In Fig.~\ref{fig:theory} we show the theoretical values for the expectation and variance of the annulus variance in Eqs.~(\ref{expsig2}) and (\ref{varvar}), compared to simulations based on a {\em noiseless} full-sky map from the best-fit {\it WMAP\/} 7-year power spectrum~\cite{Larson}, convolved with a $13.2^{\prime}$ Gaussian beam (which is close to the W-band beam size). We use {\em only} the CMB component for simplicity, since the ensemble average power spectrum (required to compute the theoretical values) is known in advance (whereas the actual W-band map also contains noise and a foreground component). This is why $\sigma^2 (R)$ in Fig.~\ref{fig:theory} is slightly below the levels in the previous figures. Note that Figs.~\ref{fig:Wband}--\ref{fig:Wtri} plot the standard deviation, $\sigma$, to enable direct comparison with GP10, while Fig.~\ref{fig:theory} plots the variance, $\sigma^2$.  This is because the above derivation does not carry over straightforwardly to the standard deviation.

 We see that the theoretical values  agree very well with the results of actual calculations of the variance for many centres.  The suppression in variance observed for small $R$ is simply due to the presence of structures in the CMB on the prefered scale of the first acoustic peak at $\sim$$1^\circ$: the smallest annuli are roughly this size, and hence the temperature is expected to be more strongly correlated, i.e.\ the variances lower, than for annuli with larger $R$.

In summary, the CMB probes the largest accessible scales, and so it is certainly worth looking for signatures which may be smoking guns for new physics (from the extensive literature on this topic, we randomly suggest~\cite{tilt}). Such signatures certainly {\it could\/} reveal information about the earliest stages of the Universe.  We have confirmed the presence of the low-variance circles in the CMB sky found by GP10.  However, we have shown that such circles are to be expected in typical realizations of the CMB sky based on the usual assumption of Gaussian random primordial fluctuations with the scale dependence which is now so well measured.  So if there are signals of extraordinarily early times buried in the CMB, they have not yet been found, and we will have to keep looking.

{\em Acknowledgments.---}%
This research was supported by the Natural Sciences and Engineering Research Council of Canada and the Canadian Space Agency.  We thank Andrei Frolov and Jonathan Benjamin for useful discussions. Use was made of the HEALPix \cite{Gorski} software package. After this study was completed we became aware of two other papers \cite{WehEri,Hajian} which come to essentially identical conclusions.

%\bibliography{bib}

\begin{thebibliography}{9}
\bibitem[{tri()}]{triangles}
\bibinfo{note}{One could instead imagine searching for other shapes, even
  special shapes which tile the sphere.}
\end{thebibliography}

\end{document}